\documentclass[acmsmall]{acmart}

\usepackage{soul}
\usepackage{tfrupee}
\usepackage[normalem]{ulem}
\usepackage{natbib}

\usepackage{xcolor}
\definecolor{control}{RGB}{0, 76, 109}
\definecolor{intervention}{RGB}{105, 150, 179}
\definecolor{white}{RGB}{255, 255, 255}
\definecolor{black}{RGB}{0, 0, 0}
\definecolor{median}{RGB}{237, 157, 66}

\newcommand{\sqboxs}{1.2ex}% the square size
\newcommand{\sqbox}[1]{\textcolor{#1}{\rule{\sqboxs}{\sqboxs}}}
\newcommand{\sqboxblack}[1]{\setlength{\fboxsep}{0pt}\fbox{\sqbox{#1}}}

\newcommand{\ed}[1]{{ %\color{blue}
{#1}}}

\newcommand{\sys}[1]{\textit{Empathosphere}}

\setcopyright{acmcopyright}
\copyrightyear{2022}
\acmYear{2022}
\acmDOI{10.1145/1122445.1122456}

\acmConference[CSCW]{Proceeedings of the ACM:CSCW}{2022}{Taipei, Taiwan}
\acmBooktitle{Proceeedings of the ACM: CSCW, 2022}
\acmPrice{15.00}
\acmISBN{978-1-4503-XXXX-X/18/06}

\begin{document}

\title[\sys{}: Improving Communication in Ad-hoc Virtual Teams]{\sys{}: Promoting Constructive Communication in Ad-hoc Virtual Teams through Perspective-taking Spaces}

\author{Pranav Khadpe}
\affiliation{%
  \institution{Carnegie Mellon University}
  \city{Pittsburgh}
  \state{Pennsylvania}
  \country{USA}}
\email{pkhadpe@cs.cmu.edu}

\author{Chinmay Kulkarni}
\affiliation{%
  \institution{Carnegie Mellon University}
  \city{Pittsburgh}
  \state{Pennsylvania}
  \country{USA}}
\email{chinmayk@cs.cmu.edu}

\author{Geoff Kaufman}
\affiliation{%
  \institution{Carnegie Mellon University}
  \city{Pittsburgh}
  \state{Pennsylvania}
  \country{USA}}
\email{gfk@cs.cmu.edu}

\renewcommand{\shortauthors}{Pranav Khadpe, Chinmay Kulkarni, Geoff Kaufman}

\begin{abstract}
When members of ad-hoc virtual teams need to collectively ideate or deliberate, they often fail to engage with each others' perspectives in a constructive manner. At best, this leads to sub-optimal outcomes and, at worst, it can cause conflicts that lead to teams not wanting to continue working together. Prior work has attempted to facilitate constructive communication by highlighting problematic communication patterns and nudging teams to alter their interaction norms. However, these approaches achieve limited success because they fail to acknowledge two social barriers: (1) it is hard to reset team norms mid-interaction, and (2) corrective nudges have limited utility unless team members believe it is safe to voice their opinion and that their opinion will be heard. This paper introduces \sys{}, a chat-embedded intervention to mitigate these barriers and foster constructive communication in teams. To mitigate the first barrier, \sys{} leverages the known benefits of ``experimental spaces'' in dampening existing norms and creating a climate conducive to change. \sys{} instantiates this ``space'' as a separate communication channel in a team's workspace. To mitigate the second barrier, \sys{} harnesses the benefits of perspective-taking to cultivate a group climate that promotes a norm of members speaking up and engaging with each other. \sys{} achieves this by orchestrating authentic socio-emotional exchanges designed to induce perspective-taking. A controlled study ($N=110$) compared \sys{} to an alternate intervention strategy of prompting teams to reflect on their team experience. We found that \sys{} led to higher work satisfaction, encouraged more open communication and feedback within teams, and boosted teams' desire to continue working together. This work demonstrates that ``experimental spaces,'' particularly those that integrate methods of encouraging perspective-taking, can be a powerful means of improving communication in virtual teams.
\end{abstract}




\maketitle

\section{Introduction}
 \begin{figure}[t]
    \centering
    \includegraphics[width=\textwidth]{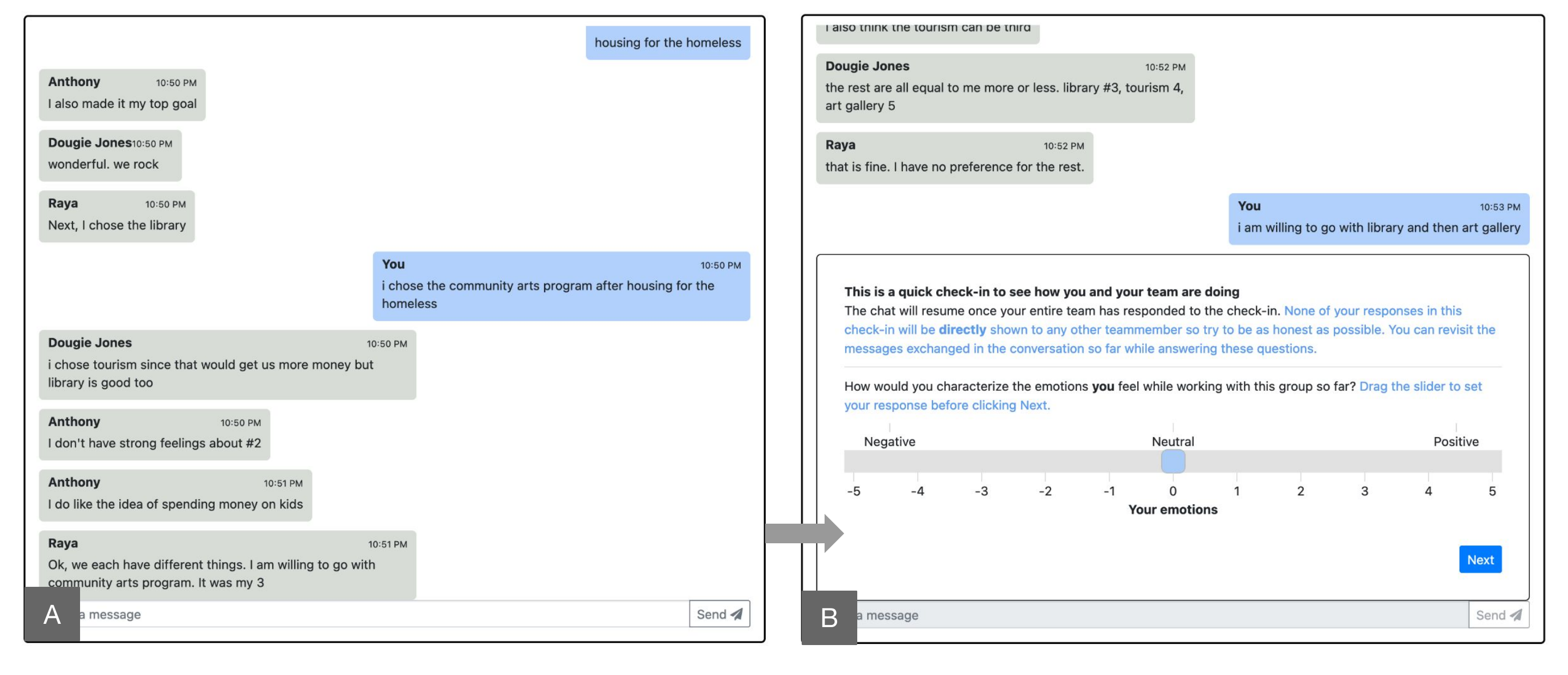}
    \caption{(A) Virtual teams often co-ordinate and communicate over chat. \sys{} helps teams establish positive interaction norms that promote sharing and engaging with each other's perspectives. (B) When triggered, \sys{ } appears as a widget in the chat interface and disables the chat while teams carry out a perspective-taking exercise. The perspective-taking exercise in \sys{} supports making team members more attuned to each others' emotions. It does this by prompting members to reflect on how others might be feeling and by providing them with socio-metric feedback on group climate and feedback on their perceptiveness towards their teammates emotions. We find that \sys{} improves work satisfaction and encourages more open communication and feedback within the team.}
    \label{fig:space}
\end{figure}

Online platforms empower people to come together, collaborate, and contribute towards common goals, often in an ad-hoc fashion - in effect forming ``virtual teams.'' Such ad-hoc virtual teams- often composed of strangers brought together by a shared purpose- have demonstrated the potential to accomplish complex tasks including authoring open-source software~\cite{sholler2019ten}, authoring open-access encyclopedia articles~\cite{bryant2005becoming}, interface prototyping~\cite{Valentine2017-qu}, collective design~\cite{Salehi2017-xo}, and even research~\cite{vaish2017crowd}. Virtual teams have indeed achieved complex goals,  but only when supported by tools and processes that decompose these goals into tasks, workflows, or well defined roles, ultimately dividing labor into tasks that \textit{individual} workers can accomplish. When virtual teams engage in tasks that cannot be decomposed, where team members need to \textit{collectively} ideate, engage with each others' perspectives, and make decisions, they often fail\ed{~\cite{olson2000distance}}: at best, their outcomes are suboptimal~\cite{bjelland2008inside}; at worst they struggle to manage inter-team conflict~\cite{kittur2007he}, which can end in team members refusing to continue collaborating~\cite{Whiting2019-wl}. Therefore, despite their promise, virtual teams are currently limited in their ability to collaborate effectively.

A possible response to these challenges of ad-hoc teams is to design systems that make disagreements and conflicts between team members less likely, so teams are more apt to succeed. Unfortunately, disagreements are often the result of diverse perspectives and ideas of individual members~\cite{Sessa1996-ba}, and so eliminating disagreements may well eliminate these diverse contributions as well. Prior work shows that disagreement is in fact a double-edged sword: when managed well, open communication of disagreement and alternate ideas can improve collaboration~\cite{Menold2019-wc, Morrison2000-uu, franco1995anatomy}. When managed poorly, disagreement can spiral out of control, triggering interpersonal conflict~\cite{kittur2007he, Whiting2019-wl}. 

Thus, when teams need to accomplish tasks that benefit from the diverse perspectives of members and require collective decision-making, surfacing and managing team friction is crucial. Specifically,  members need to be able to communicate opinions openly but without conflicts of opinion impairing teams. While all teams struggle to strike this balance, the lack of richness of computer-mediated communication compared to face-to-face communication exacerbates this problem for virtual teams. Without such balance, virtual teams are prone either to social loafing where team members avoid speaking up and voicing their opinions~\cite{kim2021moderator}, or situations where team members ignore other members' perspectives, triggering conflict and threatening the team's performance and sustainability~\cite{hinds2003out}.

To help members of virtual teams engage constructively with each others' perspectives, this paper introduces \sys{}, a chat-embedded intervention that attempts to mitigate some of the social and psychological barriers to open team communication. While the artifacts that ad-hoc virtual teams work on, appear on one set of platforms (e.g. Github in the case of open source teams), the work of discussion and deliberation often occurs through real-time discussion channels like chat~\cite{Salehi2018-vt, Zhou2018-mk}: open source teams have Discord servers~\cite{discord}, Wikipedians use IRC~\cite{wikiirc}, and some teams of crowdworkers use Slack~\cite{Salehi2018-vt, Whiting2019-wl}. The primary role of chatrooms in facilitating team deliberation make them an important setting for intervention.

\sys{} identifies and addresses two potential barriers to healthy communication in virtual teams based on literature in organizational psychology. The first barrier is that groups require a climate of \textit{safety} and \textit{efficacy} for members to speak up- team members speak up only when they feel 
% their opinion matters (\textit{efficacy}) , 
that it is safe to voice their opinion (\textit{safety}) and that their opinion will be taken into account (\textit{efficacy})~\cite{morrison2011speaking}. The second barrier is that that team norms can be hard to reset mid-interaction~\cite{marks2001temporally, Whiting2020-pq}. \ed{In response, \sys{} induces perspective-taking as a useful tool to maintain the \textit{safety} and \textit{efficacy} essential for team members to engage with each other. To transform established norms and cultivate \textit{safety} and \textit{efficacy} by inducing perspective-taking, \sys{} leverages the power of ``experimental spaces''~\cite{zietsma2010institutional, furnari2014interstitial} which are social settings separated by physical, temporal or symbolic boundaries from everyday work that diminish the salience of existing interaction norms and catalyze the emergence of new dynamics~\cite{Lee2020-zn, bucher2016interplay, furnari2014interstitial, zietsma2010institutional}.}

\sys{} instantiates a norm-diminished "experimental space" through a temporally and spatially separated channel within a chat-based collaboration environment (Fig. \ref{fig:space}). Within \sys{}, we orchestrate an exercise aimed at inducing perspective-taking. \ed{By uniting the benefits of experimental spaces and perspective-taking, we find that \sys{} improves work satisfaction, encourages more open communication and feedback within teams, and boosts team members' desire to continue working together, i.e. team viability.} 

In doing so, \sys{} shifts the focus from easily observable team communication patterns to the subtle (but vitally important) social and psychological barriers preventing team members from changing these patterns. For instance, prior work has highlighted problematic communication patterns by providing sociometric feedback~\cite{Leshed2007-up,Kim2012-jv,Leshed2009-eq,Tausczik2013-br} and by making explicit recommendations for behavior change ~\cite{Tausczik2013-br, kim2021moderator}. These systems assume that highlighting problematic patterns and suggesting changes is sufficient to transform behaviors. However, because they ignore the social and psychological barriers preventing team members from speaking up, they often have adverse outcomes: a system that displayed group-level agreement led to team members expressing agreement with a majority opinion even if they did not agree with it, in attempts to improve displayed agreement~\cite{Leshed2009-eq}. 

To test if \sys{} is effective, we conducted a controlled-study ($N=110$ in 24 teams) where we compared the benefits of \sys{} against an alternative intervention in which participants simply reflected on their team experience. In both conditions, participants recruited from Amazon Mechanical Turk worked synchronously on a decision-making task, in teams of 4-6 via a chat-based collaboration system. 

We found that participants in the \sys{} condition were significantly more satisfied with their team's solution and they judged their teams to be significantly more viable. Furthermore, we find that participants in the \sys{} condition were approximately 25.4\% more likely to be willing to provide feedback to teammates than in the control condition, and a larger fraction of participants  were willing to receive feedback from teammates. Analyzing linguistic patterns exhibited  by teams during deliberation in the two conditions, we find that \sys{} increased teams' usage of second-person pronouns, suggesting that the conversation was more other-focused: members engaged with what other team members were saying and drew them into the conversation. \sys{} also led to an increase in informality and netspeak in the teams' chatlogs suggesting that team members formed stronger social bonds.

This work contributes embedded interventions that use ``experimental spaces'' coupled with protocols to enhance perspective-taking as a powerful tool in establishing positive interaction norms in virtual teams. This opens up possibilities to use such spaces to inspire counternormative behaviors that can improve group climate and even, amplify minority voices in teams. With \sys{}, we suggest that mitigating the social and psychological barriers to changing interaction norms is a crucial step in actually transforming them. 
\section{Related Work}
This research contributes to a growing body of work that aims to improve collaboration experiences of computer-mediated teams, tying together research in organizational behavior with research in HCI. We start by reviewing existing HCI research on software support for teams and highlight the need for systems that foster positive interaction norms. We then characterize interaction patterns of performant and sustainable teams by drawing  on research in organizational behavior. Finally, we outline potential approaches for designing effective interventions to cultivate these constructive interaction patterns in teams.     

\subsection{Beyond team assembly and task support: designing systems to nurture teams}
To improve outcomes of computer-mediated teams, computer-supported collaborative work (CSCW) researchers have developed systems that aim to assemble the ``right team'' for the task. Prior work has shown that teams composed of members with higher skill diversity~\cite{horwitz2007effects}, demographic diversity~\cite{bear2011role}, and differing personality types~\cite{lykourentzou2016personality} perform better and are more satisfied. Consequently, researchers have developed systems that optimize for these criteria and assemble teams with the most promise for success. Researchers have also built ``team-dating'' systems to improve collaboration. These systems allow people to experience work with different potential collaborators before team formation~\cite{Lykourentzou2017-yz}. Other systems rotate members strategically to promote mixing of ideas~\cite{Salehi2018-vt}. To allow teams to cope with evolving work, researchers have also designed systems to on-board team members with the right expertise in near real-time~\cite{Valentine2017-qu, Retelny2014-gk}. However, while these systems have demonstrated success in identifying good collaborators, team composition is naturally constrained by availability of talent. How can we help \textit{any} team achieve their full potential and work effectively?

Once teams are formed, one way to help them work effectively is to provide support for task completion and coordination of effort. CSCW researchers have designed several systems to support team tasks. This includes systems to create and manage workflows~\cite{Valentine2017-qu, rahman2020mixtape}, systems to structure interaction between collaborators~\cite{park2021facilitating, rahman2020mixtape}. These systems decompose work into tasks that can be handled individually by members. As such, they reduce the amount of task planning that teams need to do and lower the need for members to communicate with each other. Consequently, when tasks can be decomposed into independent fragments and members' roles can be well-defined, these systems allow virtual teams to thrive. However, when tasks cannot be completely compartmentalized, and members need to come co-ideate, deliberate, and negotiate, communication becomes crucial again, and computer-mediated teams are once again vulnerable to conflict and misunderstanding~\cite{Whiting2019-wl, Whiting2020-pq}. For instance, one prior study showed that even when task-support interventions improved the performance of participant teams on average, interpersonal conflict in some teams negated the positive effects of the intervention~\cite{Bradley2003-aa}.

% Healthy communication between team members is key to ensuring team viability.
Healthy team communication is foundational: teams with healthy interpersonal processes provide a fertile ground for the positive impact of task-support interventions to improve performance~\cite{Bradley2003-aa}. Without effective communication, team performance can suffer and can even lead to teams not wanting to continue working together. Furthermore, team performance alone doesn't paint a complete picture of how a collaboration is going. Even high performing teams might have toxic culture, unresolved conflict, divisive interactions, and member burnout~\cite{driskell1999does, einarsen2002bullying}. For a more holistic evaluation of a team, it is important to look beyond just \textit{team performance} and evaluate \textit{team viability}: the team's sustainability and capacity for continued success~\cite{Bell2011-rz,Cooperstein2017-cj, Whiting2019-wl}. Team viability is an emergent state that captures the potential of the team to continue performing well, the desire of the team to continue working together, and the extent to which the team can adapt to changing circumstances~\cite{Bell2011-rz}. Antecedents of team viability include perceived performance and interpersonal friction~\cite{Cooperstein2017-cj}. Consequently, viability requires that in addition to being performant, teams also have healthy communication and interpersonal processes in place~\cite{Cooperstein2017-cj}.
% Insert one sentence for how the above relates to / informs your approach

\subsection{Managing team-friction: identifying interaction patterns of viable teams}
% Interventions focused on improving interaction patterns in teams typically aim to improve positive exchanges in teams characterized by positive emotion and agreement~\cite{Tausczik2013-br}. However, research on what constitutes constructive team-communication patterns suggests that teams need to have the right amount of friction.
Research on what constitutes constructive team-communication patterns suggests that teams need to have the right amount of friction, which in turn must be managed well. Too much disagreement, unresolved conflict, and disregard for others' opinions and criticism can trigger negative emotions, causing members to get angry and distressed~\cite{Jung2015-px}. This distress can spiral out of control and lead to a loss of viability~\cite{Jung2015-px}. Too little disagreement and critical communication is also detrimental and can impair team performance~\cite{perlow2003silence, argyris1991teaching}, team satisfaction, and consequently, viability. However, for teams to cultivate and benefit from a ``just right'' amount of friction - where members communicate opinions openly but opinion conflicts don't hinder teams - teams must manage conflict so that members feel it is both \textit{safe} and \textit{effective} to voice their opinions~\cite{morrison2011speaking}.  

Team members feel \textit{safe} expressing themselves when they do not anticipate potential negative consequences of speaking up~\cite{morrison2011speaking}: they can voice their opinion without fear of being attacked or reprimanded for speaking up. However, safety is not enough; to improve outcomes, team members must also engage with others' perspectives, foster inclusivity, and affirm each other~\cite{Salehi2018-vt, Carmeli2015-qv, Hoever2012-hn}. This not only allows teams to incorporate diverse opinions of members but also signals to members that their opinions will be taken into account and acted upon. This can persuade members that voicing their opinions is an \textit{effective} way to enact change and, in turn, motivate them continue to voice their opinions. 

The diverse perspectives and unique expertise of individual members make teams potent and so, for teams to be effective, it is crucial that team-members both speak up expressing their perspectives and that they engage with others' perspectives, in a respectful manner~\cite{morrison2011speaking}. Eliciting diverse perspectives by fostering open communication of ideas, opinions, and issues can help uncover new solutions~\cite{Menold2019-wc}, improve decision making, facilitate team learning~\cite{morrison2011speaking, Morrison2000-uu}, and help identify problems. 

% A large body of work in organizational behavior has shown that positive relational dynamics are crucial for healthy functioning of teams~\cite{Lee2020-zn}. Respect, openness, connectedness, effective conflict management, and affect management can improve team creativity~\cite{Carmeli2015-qv}, resilience~\cite{Stephens2013-vx}, performance~\cite{Carmeli2015-qv,Edmondson1999-wl}, and ultimately, viability. Respectful interactions including engaging with the diverse perspectives of team members, ensuring inclusivity, and making sure people’s ideas and opinions are respected~\cite{Lee2020-zn} lead to higher cohesion and creativity~\cite{Carmeli2015-qv}. Teams with higher psychological safety and open communication place value on critical communication. Such open communication of disagreement and issues can lead to enhanced performance~\cite{Edmondson1999-wl} and decision making~\cite{Morrison2000-uu}. 

CSCW researchers have developed methods to detect whether the interaction patterns in a team are healthy~\cite{Vrzakova2019-vk,Stewart2019-cs}. They have developed models that can leverage information about facial expression, eye gaze, head movements, speech, text, physiology, and interaction to detect the presence of positive traits~\cite{Cao2021-gm,Stewart2019-cs,Vrzakova2019-vk} such as turn-taking~\cite{Jokinen2013-ai}, interactivity~\cite{Ashenfelter2007-tk}, high-levels of empathy~\cite{Ishii2018-ex}, and team cohesion~\cite{Muller2018-ed}. These models have also enabled predicting team performance and team viability from a team’s interactions~\cite{Cao2021-gm}. While this line of work has furthered our understanding of which interpersonal processes and behaviors promote team viability, cultivating these prosocial behaviors in teams remains challenging~\cite{Lee2020-zn}. Our work attempts to move from these descriptive insights towards prescriptive interventions to help teams develop positive communication patterns.  

\subsection{Fostering constructive interaction patterns in teams}
Prior work has attributed the difficulty of establishing healthy communication to the low levels of information richness, lack of social cues, context, and shared emotion in computer-mediated teams~\cite{Hinds2003-ky}. In response, researchers have developed more information-rich communication channels to improve social cues and emotional expression in these messaging systems. This includes systems to communicate affect by altering the font and animation of text~\cite{buschek2015there,lee2006using,wang2004communicating}, and via abstract visuals~\cite{viegas1999chat,tat2002visualising,tat2006crystalchat}. Researchers have also explored the design of computer-mediated interventions to nudge people towards more positive interaction patterns by providing them socio-metric feedback on group communication patterns~\cite{Leshed2007-up,Kim2012-jv,Leshed2009-eq,Tausczik2013-br}. Some tools go beyond making extant behaviors salient and also make specific recommendations on how teams should alter their practices to move them closer to desired behaviors~\cite{Tausczik2013-br, kim2021moderator}. However, these systems have only had limited success, because they assume that highlighting problematic patterns and providing information-rich communication channels is sufficient to nudge team members to express their authentic opinions and change their interaction norms. In doing so, they ignore the social and psychological barriers preventing team members from speaking up: specifically, they do nothing to alter team members' perceived \textit{safety} and \textit{efficacy} of speaking up, necessary for open and healthy team communication. This can lead to undesirable outcomes: prior work found that visualizing group-level agreement led to a form of social loafing, where team members expressed agreement with the majority opinion even if they did not agree with it with it, ultimately resulting in lower quality work~\cite{Leshed2009-eq}.

% \pranav{emphasize that norms stick} 
Transforming and shaping interaction norms to promote open and respectful communication is challenging. It requires team members to take initiative and proactively engage in open and frank communication but team members often do not speak up. They may view expressing disagreement, voicing concerns, or asking questions as risky (low levels of perceived \textit{safety}), or they may believe that their inputs will not be taken seriously~\cite{morrison2011speaking} (low levels of perceived \textit{efficacy}), pushing them to avoid disagreeing entirely. These beliefs about speaking up are shaped by the interaction history of the team~\cite{morrison2011speaking}. For newly convened teams, an uncertainty about group norms and potential consequences of speaking up could lead to low perceived \textit{safety} and discourage individuals from speaking up. Team members who do not strongly identify with their team, or who are unsatisfied with their team, might perceive low levels of \textit{efficacy} and think speaking up is futile~\cite{morrison2011speaking}. In such cases, a ``climate of silence''~\cite{Morrison2000-uu} becomes the norm, which can undermine team performance and viability~\cite{perlow2003silence, argyris1991teaching}.

\subsection{Psychological theories informing the design of our intervention}
In designing our intervention, we leverage perspective-taking processes as a useful tool to maintain the \textit{safety} and \textit{efficacy} necessary for healthy team communication. To  support the development of perspective-taking effectively, we design an environment that is conducive to the emergence of new practices. This environment is designed based on the concept of ``spaces'' (sometimes called ``counter-normative", ``experimental'', or ``transformative'' spaces) that provide a zone of \textit{safety} for experimental behaviors, which can then develop into a perspective-taking exercise.

\subsubsection{Creating an experimental space to induce perspective-taking} To suspend pre-existing interaction norms and induce perspective-taking, we draw on work that has explored the benefits of creating “spaces'' for transformative dialogue~\cite{Lee2020-zn, zietsma2010institutional}. Spaces are social settings separated by physical, temporal or symbolic boundaries from everyday work within which the salience of existing norms and interaction patterns diminishes. This creates an environment conducive to risky and counternormative behaviors that can catalyze the emergence of new dynamics~\cite{Lee2020-zn, bucher2016interplay, furnari2014interstitial, zietsma2010institutional}. Examples of such spaces include creating meetings to specifically engage in discussion about work challenges or organizing a retreat to help team members connect outside of work environments. Such “experimental spaces'' have been effective at reducing barriers to open communication in teams and consequently, fostering positive relational dynamics~\cite{Lee2020-zn}. In this paper, we explore what such a space might look like in the context of online team communication environments and how we might use such a space to induce perspective-taking towards the goal of improving team interactions.

\subsubsection{Inducing perspective-taking to maintain \textit{safety} and \textit{efficacy}}

Perspective-taking can be a useful tool to maintain perceived \textit{safety} of speaking up. By taking another’s perspective, team members are more likely to anticipate disagreement, recognizing that other people will have different views~\cite{Sessa1996-ba}. This can both reduce initial opposition to others' ideas, as well as mentally prepare individuals to handle opposition to their ideas. Both of these together contribute to \textit{safety}.

Perspective-taking encompasses considering others’ evaluative standards which can lead to better evaluation of their ideas and more constructive feedback~\cite{Hoever2012-hn}. This can help team members feel heard and contribute to improved perceptions of \textit{efficacy}. Trying to understand someone else’s perspective can also trigger clarifying communication that can ultimately facilitate better understanding of everyone’s ideas~\cite{Hoever2012-hn}. Perspective-taking can lead to a cognitive reframing that leads to better integration of others’ ideas~\cite{Hargadon2006-ri}. Taken together, this suggests that perspective-taking can help teams take everyone's opinions into account and make everyone feel heard, both of which improve perceived \textit{efficacy} of speaking up.

\section{\sys{}}
\begin{figure}[t]
    \centering
    \includegraphics[width=\textwidth]{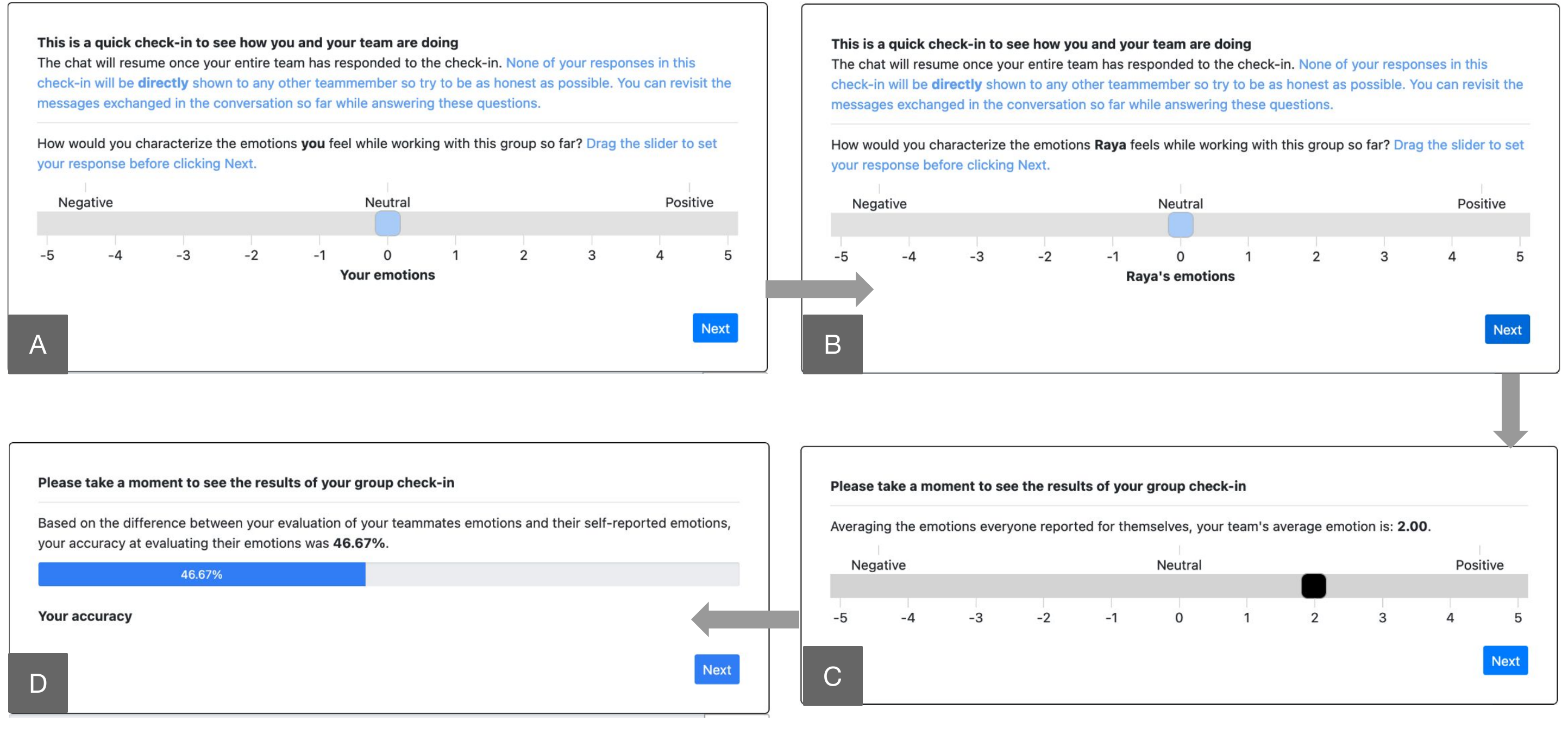}
    \caption{\sys{} creates a separated space within a team's chat within which members engage in a perspective-taking exercise. They A) share their own socio-emotional states, B) think about the socio-emotional states of their teammates and finally, \sys{} presents them with C) a measure of the group’s cumulative socio-emotional state and D) a measure of how accurate they were at evaluating their teammates’ true socio-emotional states.}
    \label{fig:workflow}
\end{figure}

To prompt team members to engage in perspective-taking, we designed and implemented \sys{}, a ``space'' embedded within a chat-based collaboration interface within which participants engage in a perspective-taking exercise. To explore different possible designs within a completely configurable chat environment, we built a custom chat-based collaboration environment (Figure \ref{fig:space}). This was a client-server web application built using Meteor.js. \sys{} attempts to improve team communication and perspective taking by providing an experimental space that encourages members to express their authentic emotions and prompts reflection on others' emotions. In doing so, \sys{} uses the power of ``spaces'' to facilitate counternormative behaviors. When triggered, \sys{} appears as a widget in the chat interface and disables the chat while teams carry out the perspective-taking exercise. The perspective-taking exercise (Figure \ref{fig:workflow}) in \sys{} was designed by drawing on known strategies to regulate perspective-taking.

\subsection{Design of the perspective-taking exercise}
A person's ability to empathize or take-perspectives is not an immutable trait; instead it can be regulated by \textit{attention modulation}: drawing attention to others' emotions and affective states can increase perspective-taking~\cite{zaki2014empathy,curran2019understanding}. \sys{} prompts this process by asking each member to reflect on how other members might be feeling and by highlighting discrepancies between their evaluation of the group and the group's actual emotions. 

To aid reflection and discrepancy evaluation, while still maintaining the ambiguity and anonymity necessary to prompt authentic expression, \sys{} executes this process in three stages. First, \sys{} privately elicits how team members actually feel about working with their group (Figure \ref{fig:workflow}A) on a scale ranging from -5 to 5; -5 being the most negative and 5 the most positive.  Next, it asks each member, in private, to guess how each of the other members in the team might be feeling on the same scale, to nudge them to direct their attention towards others in the team (Figure \ref{fig:workflow}B). Finally, \sys{} calculates the mean of responses from the first stage to present each participant with feedback about the aggregate group climate, i.e. how positive/negative the group as a whole is feeling, without revealing individual responses (Figure \ref{fig:workflow}C). (While emotions i.e., how people feel, can be multi-faceted, we focus on a simple positive/negative model in our intervention to ease reflection and evaluation for participants.) Forcing reflection on others' emotions and presenting the average group climate are the first way \sys{} draws members' attention to the possibility that some team members might not be having a positive team experience. 

Beyond attention modulation, perspective-taking is also regulated by \textit{appraisal} or an individual's evaluation of the authenticity and intensity of others' emotions~\cite{zaki2014empathy}. Emotions expressed privately in the ``space'' are likely to be more authentic and uninhibited than those expressed in the chat. So, team members are more likely to recognize potential grievances and act on them when they are confronted with them within the space. To further aid in perspective-taking, \sys{} also presents every member with feedback on how accurate they were at guessing others' emotions (Figure \ref{fig:workflow}D) in their responses in the second stage above. \ed{Specifically, we use a representation of mean absolute error that facilitates easier interpretation. For a team member $i$, the accuracy $\mathcal{A}_i$ is calculated as below, where $G_{ji}$ is $i$'s guess about $j$'s emotional state, and $S_j$ is $j$'s self-reported emotional state:} 
\begin{equation*}
    \mathcal{A}_i = max(0, 1 - \frac{1}{5}\sum_{i,i\ne{}j}^n \frac{|G_{ji}-S_j|}{n-1})
\end{equation*}

\ed{In an earlier version of \sys{}, we presented the raw mean absolute error, however, participants in our pilot studies found it hard to interpret this number. In response, we chose to transform the mean absolute error to an accuracy metric such that the accuracy is at 100\% if a member guessed all other members' emotions perfectly, and at 0\% (in expectation) for random guesses. This measure can be negative (when disagreements exceed what can be expected by chance) but this is not useful for our reflective purposes, so we show participants 0\% accuracy at minimum.} $\mathcal{A}_i$ is displayed to team member $i$ as a percentage for easier interpretation.
% would mean the average of the differences was more than $5$, i.e. the member guessed the . 
By controlling for agreements by random chance, this measure emphasizes discrepancies between an individual's evaluation of others' emotions and their actual emotions, with the goal of drawing attention to the valence and arousal levels of others' emotions.

\section{Method}
To investigate the impacts of \sys{} on teams' expression and handling of conflicting opinions and diverse perspectives, we conducted a between-subjects study. This study was approved by our university's Institutional Review Board. 

\subsection{Participants}
Participants were recruited from Amazon Mechanical Turk (AMT). We restricted participation to workers located in the United States who had completed at least 100 tasks and had an approval rate greater than 95\%. All participants were compensated at a rate of \$15/hr with a bonus that was adjusted based on the time they spent on the experiment. Prior work on team viability has studied teams ranging in size from four to eight~\cite{Whiting2020-pq} members so we aimed to form teams of similar size in our study. We wanted to ensure that teams studied had at least four members so we split participants into groups of six to account for subsequent reduction in group size due to disconnections and dropouts. We only analyzed data from teams that had at least four members till the end of the task. If at any point the number of team members dropped below four, the task was terminated and participants were compensated for their time. In total, 38 participants were unable to complete the full study either because they dropped out themselves or because their teammates dropped out and the team strength dropped below four.

\subsection{Experimental setup}
We extended the Meteor.js collaboration application that houses \sys{}, with the TurkServer~\cite{mao2012turkserver} framework. This allowed us to recruit participants on AMT and draw them into a lobby where they waited for other participants to arrive. Once there were a sufficient number of participants, the application automatically assigned teams to the different experimental conditions and created multiple parallel worlds where different teams worked simultaneously. Before they were added to chatrooms with their teams, participants were asked to set a pseudonym for themselves to preserve privacy, allowing them to control what aspects of their identity they wanted to reveal. Teams were randomly assigned to either the \sys{} or control condition:

\textit{\sys{} Condition}: For teams in this condition, \sys{} was triggered at the midpoint of the task and they were prompted to carry out the perspective-taking exercise.

\textit{Control Condition}: Teams in the control condition were asked to take a two-minute pause and reflect on their teamwork experience individually. The specific prompt we used was: \textit{``The experiment will proceed after a brief two-minute pause. Use this time to revisit the messages exchanged in the conversation so far and reflect on how the experience of working with this group has been.''}

We chose this silent, individual reflection activity  to design a conservative control condition. Prior work suggests that a break in work can have some benefits for team members' ability to navigate conflict, irrespective of the activity they engage in during the break~\cite{lerner2015emotion, cox2016design, van2017thinking}. This research suggests that time delays can be effective at de-escalating negative emotions and encourage mindful decision making and so our control condition consists of a break with a silent individual activity. This control thus seeks to isolate the effects of processes engaged by specific activities in a counter-normative space, rather than outcomes that could be attributed to the presence of a break. 

Through our pilots, we observed that the perspective-taking exercise took around two minutes on average and so, the control group's break duration was set to this time; so that the average time spent on the task would be similar for both the control and the \sys{} group. Preserving the task time between conditions seeks to further eliminate differences in familiarity between team members as a potential confound.

\subsection{Task}
Similar to previous studies of group work~\cite{Whiting2019-wl,Whiting2020-pq} and group decision making~\cite{He2017-xw}, we asked teams in both conditions to work on a negotiation task. The task required teams to decide how to allocate funds across competing project proposals. This task was an instantiation of a ``cognitive conflict'' task in the McGrath's Task Circumplex Model~\cite{mcgrath1984groups} and as such required teams to work interdependently and resolve conflicting opinions and perspectives to arrive at a solution. 

% Cognitive conflict tasks are also known to be practically path independent~\cite{Whiting2019-wl}--- the experience of working in a team isn't very sensitive to early team interactions. This made the outcomes measured in our study robust to teams accidentally ``getting off on the wrong foot'' and the task primarily tested the teams ability to handle conflicting opinions and perspectives. 
We use the well-established ``foundation task''~\cite{watson1988using, He2017-xw} to create our specific task: groups were asked to allocate \$500,000 across five competing project proposals, each in need of \$500,000. The specific proposals were: 1) To establish a community arts program featuring art, music, and dance programs for children and adults 2) To create a tourist bureau to develop advertising and other methods of attracting tourism into the community 3) To purchase additional volumes for the community's library system 4) To establish an additional shelter for the homeless in the community 5) To purchase art for display in the community's art gallery. While the original ``foundation task''~\cite{watson1988using} had a single phase where participants decided allocations, our experiment adapts this task to use two phases (discuss and decide) as below, to better measure the effects of perspective-taking.

\subsection{Procedure}
 \begin{figure}[t]
    \centering
    \includegraphics[width=\textwidth]{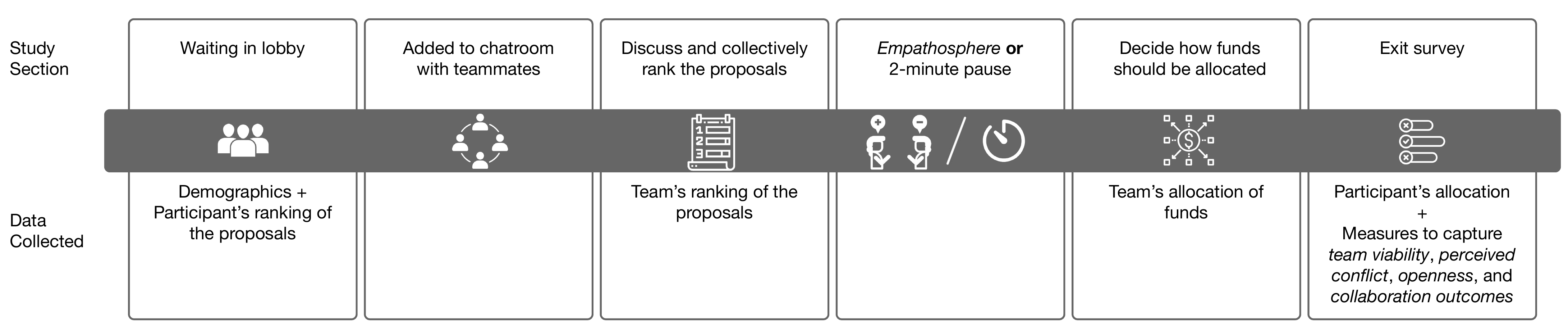}
    \caption{The study workflow and data collected in the different stages of the study.}
    \label{fig:procedure}
\end{figure}

Figure \ref{fig:procedure} illustrates the study procedure. The workflow of the study was as follows:

\subsubsection{Waiting in the lobby}
On joining the study, participants were added to a waiting room till there were a sufficient number of participants for the study. Before they were assigned to groups and the task began, each participant was presented the proposals and asked to rank them in terms of their relative merit. They were asked to evaluate the projects based on their own beliefs and values. During this time, we also collected demographic information of the participants.

\textit{Data collected:} Demographic Information, each participant's ranking of the proposals.

\subsubsection{Added to chatroom with teammates}
As soon as there were a sufficient number of participants, they were split into teams of six participants each and were added to a chatroom with their team members. Participants were asked to pick a pseudonym before they began to interact through the chat-based interface. The task was split into two phases: \textit{discuss} and \textit{decide}. 

\subsubsection{Discuss and collectively rank proposals}
In the \textit{discuss} phase, the teams were asked to weigh the pros and cons of each proposal and rank them in terms of their relative merit collaboratively. Participants were asked to advocate for projects that aligned with their personal values. At the end of this phase, we asked teams if they were able to agree on a ranking of the proposals and if so, we ask them to submit their collective ranking. Each team had nine minutes for this phase.

\textit{Data collected:} Each team's ranking of the proposals, chatlogs.

\subsubsection{\sys{} or 2-minute reflective pause}
Following this, teams in the \sys{} condition carried out the perspective-taking exercise while teams in the control condition were asked to pause for two minutes and reflect on how their experience of working in the team had been thus far. 

\subsubsection{Decide how funds should be allocated}
After this, the teams moved on to the \textit{decide} phase. Teams were informed that they had to allocate \$500,000 across the different projects. Each project was in need of \$500,000, and the more money a project got, the more likely it was to succeed. The rankings decided in the \textit{discuss} phase were meant to guide the \textit{decide} phase, but teams were told that their final allocations need not reflect their rankings from the \textit{discuss} phase. At the end of the task, we asked each team to enter their decision on how the \$500,000 should be allocated following which the participants were directed to an exit survey. Each team had nine minutes for this phase.

\textit{Data collected:} Each team's allocation of funds, chatlogs.

\subsubsection{Exit survey}
The exit survey included questions corresponding to the measures described in \ref{measures}. Finally, the exit survey also asked each participant how they would have allocated the \$500,000 themselves.

\textit{Data collected:} Each participant's allocation of the funds, responses to likert-type and open-ended questions.

\subsection{Measures}
\label{measures}
To capture different aspects of teams, their collaboration experiences, and their collaboration outcomes, we included measures for (1) baseline disagreement, (2) team viability, (3) perceived conflict, (4) openness, (5) collaboration outcomes, and (6) conversational behavior. Forms of data captured included proposal rankings provided by participants while in the lobby, chatlogs, teams' allocation of funds, and open-ended and Likert-based questions in the exit survey. Items for quantitative measures included in the exit-survey were scored on a 5-point Likert scale.

\subsubsection{Baseline disagreement}
We computed the amount of disagreement between members in a team using each participant's initial ranking of proposals as entered by them while in the lobby. For every team, we computed the Spearman footrule distance~\cite{spearman1906footrule} between the rank vectors of team members in a pairwise fashion. For every pair, the Spearman footrule distance provides a proxy for disagreement between the two members. Averaging this pairwise disagreement across all possible pairs in a team, we obtained a measure of team-level disagreement.

\subsubsection{Team viability} To measure participants' desire to continue collaborating with their teammates, we measured team viability using a three-item scale ($\alpha = 0.76$): \textit{``Most of the members of this team would welcome the opportunity to work as a group again in the future''}
\textit{``As a team this work group shows signs of falling apart.''}
\textit{``The members of this team could work together for a long time.''} The items were selected from an item pool developed to measure viability~\cite{Cooperstein2017-cj} and have been used in other studies to measure team viability~\cite{Whiting2019-wl,Whiting2020-pq}. 
 To elicit honest responses, we told participants that we might use their responses to decide whether to team them up with the same or different people if we ran subsequent experiments.

\subsubsection{Measures to capture perceived conflict}
To understand difference in perceptions of conflict across conditions, we measured perceived conflict on task-related issues- \textit{task conflict} as well as perceived interpersonal tension- \textit{relationship conflict}. 
\begin{itemize}
    \item \textit{Task conflict} was measured using a two-item scale~\cite{jehn2001dynamic} ($\alpha = 0.79$) where items were: \textit{``there was a lot of conflict of ideas in our group''}, and \textit{``my team had frequent disagreements relating to the task we were assigned.''}
    \item  \textit{Relationship conflict} was also a two-item scale~\cite{jehn2001dynamic} ($\alpha = 0.86$). Items included: \textit{``people in my team often got angry while working together.''}, and \textit{``there was a lot of relationship tension in my group.''}
\end{itemize}

\subsubsection{Measures to capture openness} Openness is characterized as frank communication of issues and feelings about both task-related and personal matters~\cite{Lee2020-zn}. To further understand differences in openness in teams across the two conditions, we measure team members' desire to give each other feedback and their receptiveness to feedback. We also include an open-ended question to probe for openness. 

Quantitative measures:

\begin{itemize}
    \item To measure participants' \textit{willingness to give feedback}, we ask them if they would be willing to give feedback to other members of this group on their teamwork practices. To elicit more honest responses, we added that giving feedback was optional, and that if they were willing, we might follow up later to collect their feedback.
    \item To measure participants' \textit{willingness to receive feedback}, we ask if they would be willing to receive feedback from other members of their team on their teamwork practices. 
\end{itemize}

Open-ended question:
\begin{itemize}
    \item We included an open-ended question asking participants if they would characterize the conversation in their group as open or guarded and asked them to explain their characterization.
\end{itemize}

\subsubsection{Measures to capture collaboration outcomes} As there is no clear performance measure for cognitive conflict tasks~\cite{Whiting2019-wl}, we measure collaborative outcomes in terms of \textit{satisfaction with solution} and the degree of  \textit{compromise}.
\begin{itemize}
    \item \textit{satisfaction with solution} captures the extent to which individual team members are satisfied with the team's final allocation of funds (survey item- \textit{``I am satisfied with my team's final solution''}). 
    \item To measure \textit{compromise} in a team, we measured the differences between participants' individual allocation of funds as provided by them in the exit survey, and the allocations that were decided by the team. For each team, we calculated the compromise measure as the mean divergence of individuals' allocations from the group's allocation. (This is calculated as the arithmetic mean of root-mean-square-of-differences between each members' allocation vector and the team's allocation vector.) This measure captures the extent to which the group's final decision aligns with individual members opinions\footnote{We also tried other measures such as the mean absolute differences between individual and group allocations. We use our current measure because it is less susceptible to outlier inputs. Results are consistent with other measures we tried.}. A higher average divergence suggests that members did not agree with the final consensus of the group--- the final decision did not involve a fair compromise. 
\end{itemize}

% \subsubsection{Team experience}
% We included other Likert-type scale questions and open-ended questions to capture a more complete picture of participants' experiences working with their teams. Statements for Likert-type questions attempted to capture \textit{respect} (\textit{``My ideas and opinions were respected by other group members''}) and \textit{satisfaction with solution} (\textit{``I am satisfied with my team's final solution''}).

\subsubsection{Conversational Behavior} 
To capture changes in the conversational behavior induced by \sys{}, we compared shifts in LIWC indicators across the two conditions. We also included two open-ended questions in the exit survey.

Quantitative measures:
\begin{itemize}
    \item \textit{Changes in LIWC indicators in the two conditions}
To analyze differences in teams' chatlog in the two conditions, we use the popular linguistic dictionary Linguistic Inquiry and Word Count, known as LIWC~\cite{pennebaker2001linguistic}. LIWC contains words bucketed into 125 psychometric categories including categories such as `first-person pronouns', `second-person pronouns', `positive emotion', `negative emotion', `informality' etc. This allows us to analyze a given text along these 125 dimensions. For every message, we compute normalized frequency counts for the LIWC categories, i.e. the number of times words from the category were present in the message divided by the total number of words in the message. To obtain psychometric measures that are independent of the length of the chatlog, we then average these normalized frequency counts across all messages in a chatlog. Therefore, the mean LIWC indicator for \textit{positive emotion} for a chatlog measures \textit{on average} how positive messages in the conversation were. 
\end{itemize}

Open-ended question:
\begin{itemize}
    \item We included an open-ended question to capture perceived changes in conversational behavior: ``How did you engage with the group in the second stage?''
\end{itemize}

% How did you engage with the group in the second stage/
% How did this compare with your experience in the first stage?

\section{Results}
A total of $N = 110$ participants completed the experiment across 24 teams with 4-6 members each. We conducted our analyses on these 24 teams which included 11 teams in the control condition and 13 teams in the intervention condition. 59\% of the participants were male and the average age of participants was 38 years ($\sigma = 10.4$). 74.5\% participants identified as White, non-Hispanic, 10.9\% identified as Asian or Pacific Islander, 7.3\% as Black, and 2.7\% as Hispanic. 21.8\% reported having a masters degree, 40\% a bachelors degree, and 36.4\% a secondary education. Figure \ref{fig:baselines}B shows a breakdown of demographics by experiment condition.

\subsection{Baseline disagreement}
\begin{figure}[t]
    \centering
    \includegraphics[width=\textwidth]{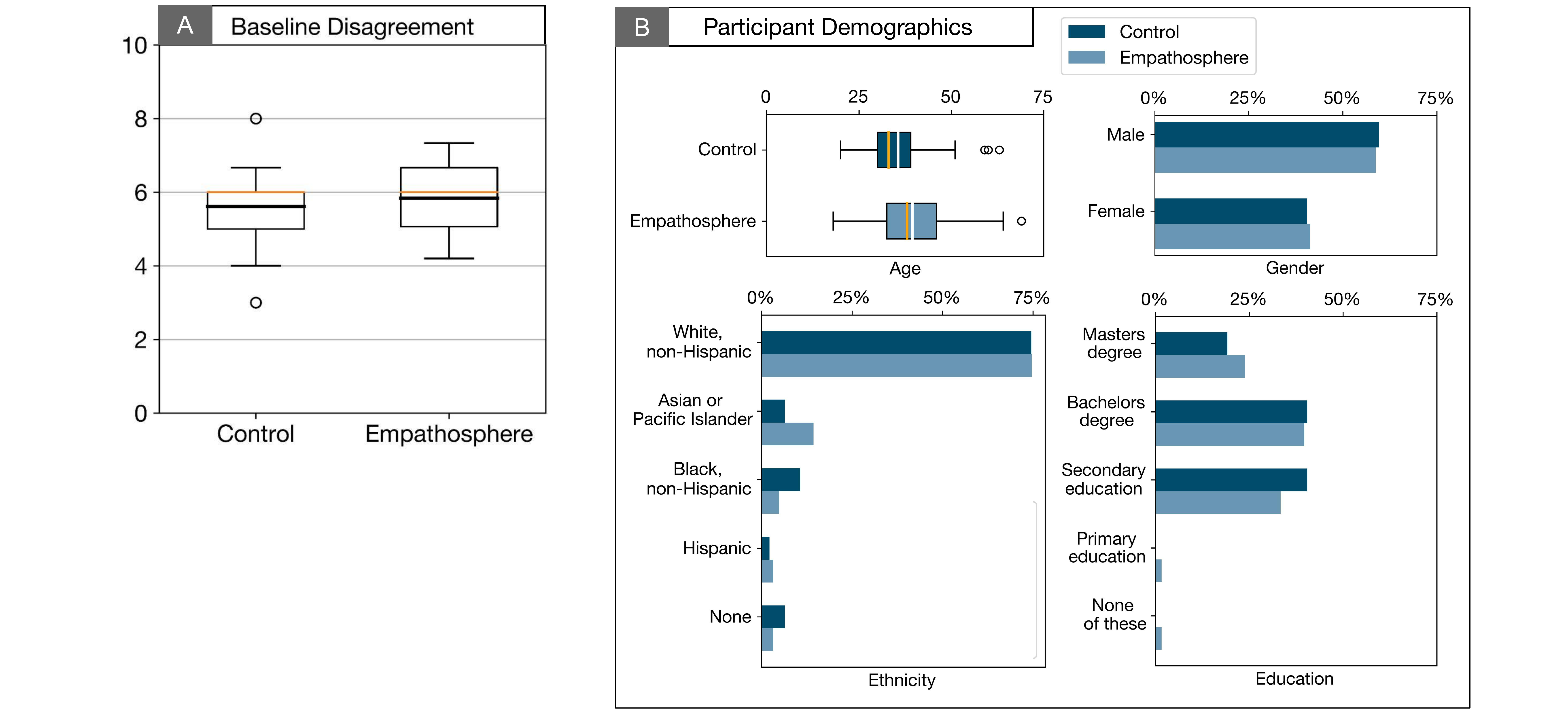}
    \caption{A) The baseline disagreement in teams in the two conditions where disagreement in a team is measured as the average of the Spearman footrule distance between the ranking of proposals across all possible pairs of members in that team. Median disagreement in both conditions is indicated by the orange marker and the mean in black. B) Demographic information of participants in our study across the control group (\sqboxblack{control}) and the intervention group (\sqboxblack{intervention}). Median age is indicated by the orange marker and mean in white.}
    \label{fig:baselines}
\end{figure}
To check whether team composition varied significantly across the two conditions, we compared the baseline disagreement in teams in the two conditions (Figure \ref{fig:baselines}A). We found no significant difference between disagreement in the control condition teams ($\mu = 5.6, \sigma = 1.27$) and intervention condition groups ($\mu = 5.83, \sigma = 1.01$) via a t-test: $t(109)=-0.45, p=0.65$. Hedges' $ g=0.197$.

\begin{figure}[t]
    \centering
    \includegraphics[width=\textwidth]{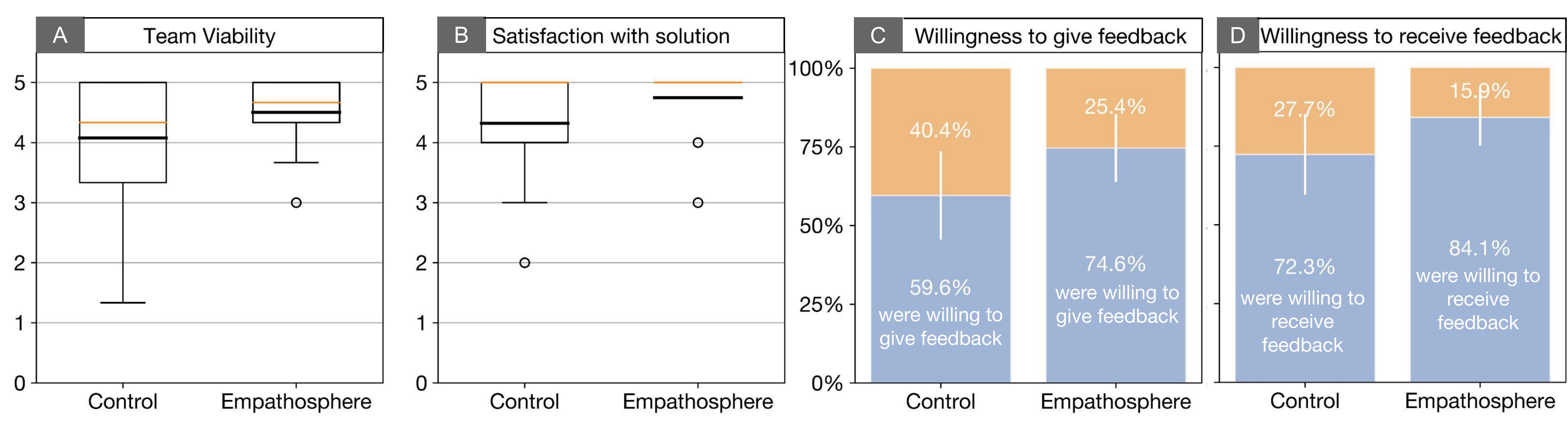}
    \caption{A) Compared to the control condition, \sys{} led to significantly higher team viability. \ed{Median score is indicated by the orange marker and the mean by the black marker.} B) Participants in the \sys{} condition expressed significantly higher satisfaction with their teams' solution than participants in the control condition. \ed{Median score is indicated by the orange marker and the mean by the black marker.} C) Participants in the \sys{} condition were more likely to give their teammates feedback. \ed{The chart shows the proportion of participants that were willing and unwilling to give feedback to other team members, with 95\% CI at the boundary.} D) Participants in the \sys{} condition were also more open to receiving feedback from their teammates. \ed{The chart shows the proportion of participants that were willing and unwilling to receive feedback from other team members, with 95\% CI at the boundary.}}
    \label{fig:results}
\end{figure}

\subsection{Team Viability}

\begin{table}[]
\resizebox{0.9\columnwidth}{!}{%
\begin{tabular}{lllllllllll}
 &  & \multicolumn{9}{c}{\textit{Outcome Variables}}                                                \\ \cline{2-11} 
                                                &  & \multicolumn{4}{c}{Team Viability}        &  & \multicolumn{4}{c}{Satisfaction with solution} \\ \hline
\textit{Fixed Effects}                          &  & Coeff.  & SE   & $z$     & $p$                &  & Coeff.   & SE     & $z$      & $p$                 \\
(Intercept)                                     &  & 4.02*** & 0.17 & 23.43 & \textless{}0.001 &  & 4.29***    & 0.14   & 30.48  & \textless{}0.001  \\
Condition                                       &  & 0.49*   & 0.23 & 2.13  & 0.033            &  & 0.45*    & 0.19   & 2.35   & 0.018             \\
Disagreement                                    &  & -0.35*  & 0.15 & -2.34 & 0.019            &  & -0.11    & 0.13   & -0.89  & 0.372             \\
ConditionXDisagreement                          &  & 0.29    & 0.23 & 1.31  & 0.191            &  & 0.08     & 0.58   & 0.41   & 0.684             \\ \hline
\textit{Random Effects}                         &  & Var.    & SE   &       &                  &  & Var.     & SE     &        &                   \\
Team                                            &  & 0.21    & 0.17 &       &                  &  & 0.09     & 0.10   &        &                   \\ \hline
\multicolumn{11}{c}{Note: $^\ast p<0.05$; $^{**} p<0.01$; $^{\ast\ast\ast} p<0.001$}                                                             
\end{tabular}
}
\caption{Results of mixed effect linear regression analyzing the impact of experiment condition and disagreement within the group on measures of \textit{team viability} and \textit{satisfaction with solution}. The condition had a significant effect on both measures with participants in the \sys{} condition expressing higher viability and satisfaction with solution.}
\label{tab:table1}
\end{table}

\textit{\sys{} improved team viability.}
Participants in the \sys{} condition scored team viability to be higher ($\mu = 4.50, \sigma = 0.63$) than in the control condition ($\mu = 4.08, \sigma = 0.98$) (Figure \ref{fig:results}A). We fit a mixed effects linear regression model with the experiment condition and team disagreement as fixed effects, their interaction, team grouping as a random effect, and team viability scores as the outcome variable. Results in Table \ref{tab:table1}. We observe a significant effect of both the condition \ed{($\beta=0.49;\textsc{ } 95\% \textsc{ CI} = 0.06\textsc{,}0.93;\textsc{ }  p<0.05$)} and disagreement \ed{($\beta=-0.35;\textsc{ } 95\% \textsc{ CI} = -0.63\textsc{,}-0.07;\textsc{ }  p<0.05$)} on team viability, with no significant interaction effects. Teams with lower initial  disagreement had higher viability, suggesting that teams with lower opinion diversity, and therefore inherently lower potential for conflict, tend to exhibit higher viability. Meanwhile, the effect of \sys{} on team viability shows that prompting perspective-taking is a promising approach to improving viability of all teams, regardless of the degree of diversity in team members opinions.

\subsection{Perceived Conflict}

\begin{table}[]
\resizebox{0.9\columnwidth}{!}{%
\begin{tabular}{lllllllllll}
&  & \multicolumn{9}{c}{\textit{Outcome Variables}}                                           \\ \cline{2-11} 
                                                &  & \multicolumn{4}{c}{Task Conflict}         &  & \multicolumn{4}{c}{Relationship Conflict} \\ \hline
\textit{Fixed Effects}                          &  & Coeff.  & SE   & $z$     & $p$                &  & Coeff.  & SE   & $z$     & $p$                \\
(Intercept)                                     &  & 2.19*** & 0.21 & 10.64 & \textless{}0.001 &  & 1.55*** & 0.17 & 9.28  & \textless{}0.001 \\
Condition                                       &  & -0.36   & 0.27 & -1.28 & 0.199            &  & -0.15   & 0.23 & -0.66 & 0.512            \\
Disagreement                                    &  & 0.31    & 0.18 & 1.73  & 0.084            &  & 0.11    & 0.15 & 0.77  & 0.438            \\
ConditionXDisagreement                          &  & -0.04   & 0.27 & -0.13 & 0.896            &  & 0.10    & 0.22 & 0.46  & 0.644            \\ \hline
\textit{Random Effects}                         &  & Var.    & SE   &       &                  &  & Var.    & SE   &       &                  \\
Team                                            &  & 0.29    & 0.19 &       &                  &  & 0.19    & 0.15 &       &                  \\ \hline
\multicolumn{11}{c}{Note: $^\ast p<0.05$; $^{**} p<0.01$; $^{\ast\ast\ast} p<0.001$}  
\end{tabular}
}
\caption{Results of mixed effect linear regression analyzing the impact of experiment condition and disagreement on \textit{perceived task conflict}, and \textit{perceived relationship conflict} showing the absence of a significant relationship between the condition and either variables.}
\label{tab:table2}
\end{table}

\textit{We did not observe differences in perceived task or relationship conflict across the conditions}. Following the same analysis strategy as before, we fit mixed effects linear regression models with perceived task conflict and perceived relationship conflict as the outcome variables and found no significant effects of condition, disagreement, and their interaction on either outcome variable (Table \ref{tab:table2}). There was a marginally significant effect of initial disagreement on the perceived task conflict \ed{($\beta=0.31;\textsc{ } 95\% \textsc{ CI} = -0.02\textsc{,}0.66;\textsc{ }  p=0.08$)} with higher initial disagreement leading to higher perceptions of task conflict, which is in line with expectations. However, the \sys{} condition did not yield significant differences in perceived task or relationship conflict suggesting that prompted perspective-taking might not change perceptions of conflict or trigger negative emotional responses to conflict.

\subsection{Openness}

\begin{table}[t]
\resizebox{0.9\columnwidth}{!}{%
\begin{tabular}{lllllllllll}
&  & \multicolumn{9}{c}{\textit{Outcome Variables}}                                                          \\ \cline{2-11} 
                                                &  & \multicolumn{4}{c}{Willingness to give feedback} &  & \multicolumn{4}{c}{Willingness to receive feedback} \\ \hline
\textit{Fixed Effects}                          &  & Coeff.      & SE        & $z$          & $p$         &  & Coeff.       & SE         & $z$          & $p$          \\
(Intercept)                                     &  & 0.27        & 0.31      & 0.90       & 0.366     &  & 0.93*        & 0.38       & 2.45       & 0.014      \\
Condition                                       &  & 0.78        & 0.43      & 1.83       & 0.067     &  & 0.95         & 0.56       & 1.68       & 0.092      \\
Disagreement                                    &  & -0.33       & 0.29      & -1.15      & 0.252     &  & -0.62        & 0.37       & -1.65      & 0.098      \\
Condition$\times$Disagreement                          &  & 0.55        & 0.44      & 1.26       & 0.209     &  & 0.15         & 0.58       & 0.26       & 0.793      \\ \hline
\textit{Random Effects}                         &  & Var.        & SE        &            &           &  & Var.         & SE         &            &            \\
Team                                            &  & 0           & 0         &            &           &  & 0.13         & 0.37       &            &            \\ \hline
\multicolumn{11}{c}{Note: $^\ast p<0.05$; $^{**} p<0.01$; $^{\ast\ast\ast} p<0.001$}                                                                        
\end{tabular}%
}
\caption{Results of mixed effect logistic regression analyzing the impact of experiment condition and disagreement on participants' \textit{willingness to give feedback} to teammates and their \textit{willingness to receive feedback} from teammates. The condition had a marginally significant effect on both measures with participants in the \sys{} condition expressing higher willingness to give and receive feedback.}
\label{tab:table3}
\end{table}

% \subsubsection{\sys{} improved willingness to give and receive feedback but we did not observe a change in psychological safety}
\subsubsection{\sys{} improved willingness to give and receive feedback}
Since \textit{willingness to give feedback} and \textit{willingness to receive feedback} had binary responses, for each of them as outcome variables, we fit a mixed effects logistic regression model with the experiment condition and team disagreement as fixed effects, their interaction, team grouping as a random effect (Table \ref{tab:table3}). We found a marginally significant effect of condition on \textit{willingness to give feedback} \ed{($\beta=0.78;\textsc{ } 95\% \textsc{ CI} = -0.05\textsc{,}1.76;\textsc{ }  p=0.067$)}. Of the participants in the \sys{} condition, 74.6\%  were willing to give their teammates feedback while only 59.6\% of the participants in the control condition were willing to do the same (Figure \ref{fig:results}C). Similarly, we found a marginally significant effect of the experiment condition on \textit{willingness to receive feedback} \ed{($\beta=0.95;\textsc{ } 95\% \textsc{ CI} = -0.14\textsc{,}2.24;\textsc{ }  p=0.092$)}. 84.1\% of the participants in the \sys{} condition were willing to receive feedback from their teammates while 72.3\% of the participants in the control condition were willing to do the same (Figure \ref{fig:results}D). We also saw a marginally significant effect of disagreement \ed{($\beta=-0.62;\textsc{ } 95\% \textsc{ CI} = -1.47\textsc{,}0.09;\textsc{ }  p=0.098$)}, such that participants in teams with higher disagreement were less keen on receiving feedback from their teammates. 

% Fitting a mixed effect linear regression model for \textit{psychological safety} as outcome variable, we did not find any significant effects. Though participants in the \sys{} condition reported higher psychological safety ($\mu = 4.32, \sigma = 0.63$) than participants in the control condition ($\mu = 4.14, \sigma = 0.92$), the effect was not significant due to the small sample size.  

\subsubsection{\sys{} encouraged participants to voice disagreement while also making participants more perceptive to other team members' behaviors}
We compared participants' responses to the open-ended question on whether they thought the conversation in their group as open or guarded. Participants in the control condition mentioned how some team members chose to stay silent: ``Some people had nothing to say at all while two others were very open.'' (P5) Some other participants mentioned how there was very little opposition or open disagreement: ``It was not really as engaging as I hoped. I  had to get the ball rolling and didn't really get any conflicting opinions'' (P11), and ``It did not appear that anyone wanted to "dominate" the conversation/debate and therefore potentially yielded quicker than they would in person or make real decisions.'' (P41)

In contrast to this, participants in the intervention group reported higher levels of open disagreement: ``We never strongly attacked an idea and views were able to change. People were allowed to suggest their ideal solution and did not seem bothered by challenges.'' (P94), ``everyone brought something to the table and it was a great group'' (P108), ``I felt like everyone could voice their opinions, and no one was shot down unfairly.'' (P106) One participant noted a distinct shift in their behavior after the \sys{} exercise: ``I was guarded on the first stage and wanted to recommend tourism as the first priority but when others said homeless I agreed. I knew I would have to speak up quickly and I was less guard[ed] on round two because they all seemed like nice people would [who] wouldn't be rude.'' (P71)

Even though we did not explicitly probe for it, participants in the \sys{} condition also made more specific observations about their teammates suggesting that they were perceptive to how their teammates were behaving. One participant noted: ``I knew I should suggest amounts quickly because William would have a proposal and would be more particular, I think. He seems like a leader or someone who wants to be in charge and doesn't ask anyone else what they want probably. However he did compromise.'' (P71) Another participant mentioned: ``Kate seemed to be the one that had the most ideas that differed from the group. The other 2 people seemed to be the most in line with me.'' (P96)

\subsection{Collaboration outcomes}
\subsubsection{\sys{} improved participants' satisfaction with their teams' solutions}
Participants in the \sys{} condition reported higher satisfaction with their teams' solutions ($\mu = 4.74, \sigma = 0.50$) than participants in the control condition ($\mu = 4.31, \sigma = 0.94$) (Figure \ref{fig:results}B). We fit a mixed effects linear regression model with the experiment condition and team disagreement as fixed effects, their interaction, team grouping as a random effect, and satisfaction with solution as the outcome variable. Results in Table \ref{tab:table3}. We observed a significant effect of the experiment condition on satisfaction with solution \ed{($\beta=0.45;\textsc{ } 95\% \textsc{ CI} = 0.09\textsc{,}0.80;\textsc{ }  p<0.05$)} while there was no significant effect of disagreement. There was also no interaction effect.

\subsubsection{We did not observe significant differences in compromise across teams in the two conditions}
The difference between \textit{compromise} in teams in the control condition ($\mu = 0.075, \sigma = 0.022$) and teams in the \sys{} condition ($\mu = 0.070, \sigma = 0.036$) was not statistically significant, possibly due to small sample size (\textit{compromise} was measured at the group level, and so had fewer observations than some of the individual level measures above).

\subsection{Conversational behavior}
\label{ssub:conversation}
\subsubsection{Teams in the \sys{} condition used more second-person pronouns and exchanged more informal messages}
We compare the changes in LIWC indicators from the \textit{discuss} to \textit{decide} phase in the \sys{} and control conditions. We find that \sys{} was followed by an increase in use of second-person pronouns (you, you've y'all, u). Teams used 89\% more second-person pronouns in the \textit{decide} phase, after \sys{} ($p < 0.05$) but there was no significant difference in the usage of second person pronouns between the \textit{discuss} and \textit{decide} phase in the control condition. This indicates that \sys{} potentially shifted the conversation to be more other-focused, engaging with what other team members are saying and drawing them into the conversation. 

The intervention was also followed by an increase in informality (okay, yaas) and netspeak. Usage of informal words was 27\% higher in the \textit{decide} phase than the \textit{discuss} phase in the intervention condition ($p < 0.05$). Similarly, usage of netspeak was 281\% higher in the \textit{decide} phase than the \textit{discuss} phase for the intervention condition ($p < 0.001$). Meanwhile, in the control condition, the use of informal language decreased in the \textit{decide} phase compared to the \textit{discuss} phase by 24\%, however, this difference was only marginally significant ($p = 0.09$). Taken together, this also suggests that \sys{} has the potential to improve social bonds in teams~\cite{yuan2013understanding}.

\subsubsection{\sys{} helped foster higher comfort levels and led to team members respectfully engaging with each other} We analyzed participants' responses to the open-ended question asking how they engaged with the group in the second stage. Participants in the intervention condition noted higher comfort levels in their team: `It felt like coming back to a group of coworkers that I know well.'' (P91) They mentioned being able to voice their opinions with their team: ``People threw out ideas while others were more intent on keeping the group focused on coming up with an actual answer to the task.  I chimed in when I wanted something addressed or wanted to broach a specific idea that mattered to me'' (P85), ``I gave my thoughts and everyone listened and gave me constructive feedback'' (P107), and ``I suggested an alternative allocation of funds at one point and the group reached an amicable decision taking in everyone's vote'' (P67).

They also took efforts to accommodate each others' perspectives: ``I tried to take everyone's ideas and formulate them into a single plan. I took the numbers and tried to make them work, so as to make the plan easier to visualize. I think this helped the group to come to an agreement more quickly, because they could see the plan.'' (P81) They tried to make their teammates feel heard: ``I knew Williams' personality and he would have a suggestion and would want to be heard.'' (P71) Several participants mentioned how they attempted to strike a balance between voicing their perspectives and listening to others' opinions: ``I tried to take in their perspectives but wanted to make sure they understood me too'' (P94), ``I tried to summarize everyone's ideas as well as contribute my own'' (P96), and ``I made suggestions but also was open to what they thought  as well.'' (P93)

\subsubsection{Teams in the control condition had more polarized experiences, with some reporting either too little or too much conflict} Analyzing responses to the open-ended question by participants in the control condition, we found that participants hinted at loafing: ``I engaged with caution, trying to let some of the other members bounce ideas off of one another, but no one was really into it.'' (P11) Some tried to avoid conflict altogether: ``I gave my proposal, but it seemed like the group wanted to finish quickly so I didn't push my ideas that much.'' (P36) On the other hand, some participants in the control condition noted how their opinions were disregarded: ``I expressed my thoughts and ideas about how to distribute money but was getting into some arguments about the merits of some programs versus others. I was getting frustrated because it felt like my group was ignoring my suggestions'' (P24). Such dismissal of ideas also triggered negative emotions: ``I tried to keep the group focused. However, one person was not respectful of my ideas and made snide remarks about me being insecure. That definitely hampered our progress.'' (P25) 

\section{Discussion}
With \sys{}, we have demonstrated that focusing on the barriers to effective communication patterns - namely, insufficient or poorly calibrated perspective-taking - is an effective lever for providing technological support to ad-hoc virtual teams, as it provides the scaffolding for behavior change while still providing teams with the agency to regulate their own communication. We show how \sys{}, by prompting reflection about team members' emotional states, improves work satisfaction and encourages more open communication and feedback within the team, and boosts team viability. As such, \sys{} expands on and brings together several ideas that have been previously explored in HCI research. We situate our work alongside existing research focused on harnessing diverse perspectives of team members,  supporting constructive deliberation, inducing perspective-taking, and leveraging technology to support team development. We then discuss the limitations of our work and suggest possible directions for future exploration.  
% Researchers have explored two strategies for this: (1) allowing teams to invoke structure temporarily~\cite{kim2021moderator}, and (2) designing behavior change systems that promote positive communication behaviors through heightened social awareness~\cite{Leshed2007-up, Leshed2009-eq, hassib2017heartchat}. \sys{} shows the promise of systems at the confluence of these two strategies by invoking structure for micromoments and using those structured micromoments to foster behavior change.

\subsection{Harnessing disagreement to improve teamwork}
\sys{} attempts to help teams surface the diverse opinions of teams members and manage the resulting disagreement, towards the goal of improving collaborative outcomes. Prior work has established that well-handled conflict can help improve team creativity, identify and resolve issues, and strengthen common values~\cite{kittur2007he, Hoever2012-hn}. Previous systems have sought to manifest constructive conflict in different ways. Some have attempted to assemble teams by maximizing diversity in member perspectives~\cite{lykourentzou2016personality, horwitz2007effects}. Hive~\cite{Salehi2018-vt} brings new members into teams, at later stages, in an attempt to introduce fresh perspectives and challenge a teams' ideas. In doing so, it brings in conflicting opinions from sources external to the team. However, prior work has shown that mere existence of diverse opinions does not translate to disagreement and the positive benefits that arise from disagreement~\cite{Hoever2012-hn, Salehi2018-vt}, especially in the presence of a challenging group climate~\cite{morrison2011speaking}. We, instead, shift the focus from the \textit{source} of diverse perspectives towards the \textit{expression} of diverse perspectives. We attempt to illuminate the diverse perspectives and experiences of team members so that teams may benefit from their differences in perspectives. As such, this work serves to complement existing work: using established team assembly systems in concert with interventions to foster constructive communication has the potential for greater success than does either approach by itself. 

\subsection{Supporting deliberation while enforcing minimal structure}
Research aiming to improve group deliberation, has largely focused on the design of platforms that provide effective \textit{structure} for deliberation. For instance, ConsiderIt structures deliberations around user-curated pros/cons lists~\cite{kriplean2012supporting}. OpinionSpace structures deliberation by plotting opinions in a two-dimensional space~\cite{faridani2010opinion}. Lead Line structures discussion in accordance to a pre-authored script~\cite{farnham2000structured}. \ed{StarryThoughts supports visual exploration of diverse opinions while emphasizing the relationship between peoples' identities and opinions~\cite{kim2021starrythoughts}}. Structure is powerful but constraining- by reducing flexibility in how people can interact, it can encourage people to elaborate their reasoning~\cite{schaekermann2018resolvable}, encourage people to listen and consider others' opinions~\cite{kriplean2012supporting, faridani2010opinion, kim2021starrythoughts}, and allow people to scope out and consider the broader context of the deliberation~\cite{kriplean2012supporting, zhang2017wikum}. However, in practice, virtual teams rarely move to these dedicated platforms to make use of their \textit{structure} for deliberations; they deliberate within the \textit{flexible} communication channels they already occupy (for instance Discord~\cite{discord} or Slack~\cite{Whiting2019-wl, Salehi2018-vt}). This suggests the need to embed deliberation support in the unstructured, synchronous communication channels that teams choose to use while maintaining the \textit{flexibility} that drew teams to those channels in the first place. Researchers have explored two strategies for this: (1) \ed{allowing teams to invoke structure temporarily~\cite{kim2021moderator, kim2020bot}}, and (2) designing behavior change systems that promote positive communication behaviors through heightened social awareness~\cite{Leshed2007-up, Leshed2009-eq, hassib2017heartchat}. \sys{} explores the potential of systems at the confluence of these two strategies: when triggered, \sys{} creates structure for two minutes during which it attempts to heighten social awareness and facilitate behavior change. Thus, \sys{} shows the promise of invoking structure for micromoments and using those structured micromoments to foster behavior change.

\subsection{The tradeoff between control and authenticity in mechanisms to induce  perspective-taking}
\sys{} contributes an operationalization of strategies to induce perspective-taking, that relies on self-regulated emotional exchanges. A prominent line of prior work that explicitly attempts to foster empathy or perspective-taking, via emotional expressiveness, has explored the use of expressive biosignals as cues to cognitive or affective states~\cite{hassib2017heartchat, liu2017supporting}. The key difference between these approaches and ours is how emotional signals are sourced, represented, and interpreted. Unlike \sys{} that asked people to self-report their emotions, biosignals are directly sensed from a person's body and they convey personal and intimate information. So, biosignals make it harder for people to conceal or feign their emotions when compared to our approach. However, the fact that people can only regulate whether or not they share biosignals but not the extent of expression~\cite{curran2019understanding}, can make them hesitant to share their biosignals altogether~\cite{liu2017supporting}. To reduce the hesitation and to promote greater safety in the sharing of emotions, we consciously choose to give people control over the extent to which they want to express their emotions. While this agency may, in theory, come at the potential cost of authenticity, we also aimed to increase psychological safety by not revealing individual team members' own emotional self-reports, instead providing accuracy metrics at the aggregate level. Additionally, since biosignals require physiological sensing, they require users to possess these sensing systems- a constraint \sys{} doesn't have. In addition, biosignals as cues to emotions are inherently ambiguous, often providing more information about the extremity rather than the valence of an emotion and, thus, requiring subjective interpretation~\cite{curran2019understanding}. In contrast, we rely on a more straightforward reporting of emotional states, and provide an accuracy measure to help support users as they attempt to interpret others' emotions.  

\subsection{Designing interventions for different stages in a team's development process}
Truckman identifies four stages teams go through: forming, storming, norming, and performing- from when they are convened to when they realize their full potential~\cite{tuckman1965developmental}.
Forming is the initiation of collaboration where team members first come together and get to know each other. This is followed by the storming stage, where team members push against established boundaries, challenge each other, and sometimes even clash with each other. Norming is marked by team members moving past their differences, accepting each other, and this is when team members feel comfortable and safe in the team. Ultimately teams arrive at the performing stage where they have learned to work together and perform at their fullest potential. Yet, unlike Truckman's linear model, real-world teams are increasingly fluid~\cite{Bell2011-rz}, and membership changes frequently~\cite{Salehi2018-vt, Valentine2017-qu}. For such high-turnover teams, forming and storming stages are frequent: developing effective communication and accommodating new team members is an ongoing challenge. Each of these stages provides unique opportunities for technology-mediated protocols to support team development and adaptability. 

In the forming stages, teams can benefit from protocols such as team dating which help teams identify good collaborators (and prune undesirable ones) by allowing them to experience working with different collaborators~\cite{Lykourentzou2017-yz}. The storming stage is especially tricky to navigate. Several teams never move past it: escalating conflict and rising tensions can cause teams to fracture~\cite{Whiting2019-wl}. Part of the problem is that unregulated team interactions can lead to teams unintentionally establishing unhealthy interaction norms that are hard to alter later~\cite{Whiting2020-pq}. To help teams navigate the storming phase, previous work has explored the benefits of allowing teams to ``start over''~\cite{Whiting2020-pq}, by masking teammates' identities and repeatedly convening the same team till they adopt positive interaction norms. This works well when teams are assembled all at once, but masking identities and restarting is harder for a fluid, partially formed team. While this current study did not specifically study storming processes, \sys{} may empower teams to collectively navigate the storming phase by equipping team members to effectively surface and manage disagreements and conflict. \sys{} may provide teams with the opportunity to alter problematic norms they might have previously established and, as suggested by results in Section~\ref{ssub:conversation}, may also help establish healthy conversational norms, such as respectfully engaging with each other. Finally, perspective-taking interventions, if used in conjunction with protocols like Hive~\cite{Salehi2018-vt} could help help periodically expose teams to new ideas and fresh perspectives during the norming and performing stages.

\subsection{Limitations}
\subsubsection{Limitations of task design and collaboration duration}
Consistent with prior studies on group work~\cite{He2017-xw, Whiting2019-wl}, we use the negotiation task - an instantiation of a cognitive conflict task - to effectively simulate situations where team's must resolve conflicting viewpoints. Specifically, teams in our study worked on a cognitive conflict task within  McGrath’s Circumplex Model~\cite{mcgrath1984groups}.  While our results may be relevant to other kinds of tasks such as creative generative tasks that also surface, and benefit from, diverse perspectives, we did not study those tasks directly. 
% Thus, they could be a potential testing ground for \sys{}. \pranav{talk a bit about the structure of the task}

Furthermore, teams in our experiment only worked together for a short duration of about 20 minutes. This has two implications for our findings. First, we can not comment on the long term effects of our intervention and whether the changes in behavior persist beyond the short time frame. Secondly, prior work has shown that ad-hoc teams that expect to work for a short duration- as was the case with Amazon Mechanical Turk workers in our study- are less likely to be positively influenced by interpersonal interventions. The short time frame and anticipated future separation causes team members to focus on the completion of the task without attempting to form cohesive team norms that would only benefit the members if they were going to remain together for the performance of future task~\cite{Bradley2003-aa}. As such, our study setup could have potentially muted the effects of \sys{} because team members did not have an incentive to invest effort in forming group norms, which would have benefited from perspective-taking. 

% Additionally, due to the fact that we experiment with paid crowdworkers who are monetarily incentivized, much like previous studies on groupwork~\cite{Leshed2007-up, Leshed2009-eq, Whiting2020-pq}, our study context limits our ability to generalize to teams like open source software communities where goals are self defined and members unite towards common goals voluntarily. 

\subsubsection{We do not study teams with pre-existing relationships}
Participants in our study were recruited from Amazon Mechanical Turk and assigned to ad-hoc teams after they picked pseudonyms. Members in our study teams had not worked together before and even if they had, the pseudonyms made it hard for them to recognize this. While this served to mimic real world scenarios where teams might be meeting for the first time~\cite{Whiting2019-wl, hinds2003out}, we caution against generalizing our findings to teams where members might be familiar with each other and have preexisting relationships.
% Ad-hoc teams are convened online for software development, research, crisis mapping, and content creation as well as offline such as incident response teams, film crews, management consulting. However, despite their ad-hoc nature members of these teams might have varying levels of shared history and familiarity- some members might have pre-existing relationships while others might not. While our study suggests \sys{} can improve collaboration experiences in teams when members haven't worked together before, future work can investigate the impact of preexisting working relationships on emergent norms in teams and design intervention to help team members learn to work well with each other even if some of them have preexisting relationships.
\ed{\subsubsection{Our experimental design cannot quantify the effect of \sys{} on social perceptiveness}
Our qualitative analysis of groups’ conversational data, provides evidence of the impact of our intervention on team members’ displayed social perceptiveness in deliberation. However, since we did not include a direct measure of social perceptiveness in our study, we were unable to quantify the effect. Measures of changes to social perceptiveness that we considered- such as a measure that involves having participants draw a capital letter E on their foreheads and assessing whether it was drawn from a self-oriented or other-oriented perspective~\cite{hass1984perspective}- were ultimately deemed to be either too unnatural or obtrusive to fit within the flow of the study. Future work can explore the design and use of appropriate measures of social perceptiveness to test the effects of interventions like \sys{}.
}

\subsubsection{Our experimental design cannot measure the impact of team members' individual characteristics}
To protect privacy, we asked participants in our study to pick pseudonyms before they were added to chatrooms with their teams. However, doing so may have reduced the salience of their individual characterstics, such as gender and race in our study. This may have in turn muted impression formation and stereotyping as interfering phenomena.  At the same time, several participants chose gendered names, and so these effects were not entirely absent. Future work can explore the effects of team members' impressions of each other on their behaviors and how social capital, impression formation and stereotyping might interact with the effects of interventions designed to improve group norms.
% leading to possibly larger effects for \sys{}. 
% however, it also limits the generalizability of our findings. 
%  

% We also did not measure participants' innate social sensitivity or their inherent tendency to take perspectives. While random assignment of participants in our study attempted to eliminate this as a confound, 

% \subsubsection{Teams in our study collaborated only for short durations}

\subsection{Future Work}

\subsubsection{Interventions to boost minority voices in teams}
\sys{} demonstrates the promise of creating spaces that provide a layer of safety and anonymity for team members to express and reflect on their experiences of group climate. Such mechanisms, which encourage members to voice their emotions, issues, and disagreements, are critical for creating inclusive team dynamics. Psychological safety, for example, is critical for boosting minority opinions and voices in groups and centering the perspectives of individuals from marginalized groups. Prior work has shown that experiences with implicit bias and other forms of micro-inequities threaten the confidence and self-esteem of minority group members ~\cite{hall1982classroom, sandler1996pedagogy, flam1991still}. Even when members of the group might care about fairness and equitable participation, a lack of intentionality in enforcing inclusive norms can create a ``chilly climate'' for marginalized or underrepresented group members~\cite{isbell2012stag}. This problem is amplified by the fact that minority group members might feel hesitant and powerless to express their issues openly for fear of judgment and further alienation~\cite{tatum2013classroom, hill2010so}. Ultimately, this can cause minority members to leave such groups~\cite{beasley2012they, holleran2011talking}. Future work can explore the design of interventions that build on \sys{} and have particularly powerful benefits for increasing equity of participation and inclusion for diverse teams.  

\subsubsection{Expanding the notion of ``experimental spaces'' and exploring the role of virtual spaces in transforming team norms}
Researchers using ``experimental spaces'' to transform team norms have primarily done so by moving team interaction to different social settings- demarcated spaces (contributing to their sense of \textit{space}) with well defined rules of interaction (contributing to their sense of \textit{place}). However, most of this work has explored the role of ``experimental spaces'', situated in the real world~\cite{zietsma2010institutional}. \sys{} explores the alternate possibility of creating spaces in the \textit{virtual} world, where teams convene increasingly often. However, exciting opportunities might also lie at the intersection of these two approaches in hybrid spaces- virtual spaces overlayed on physical realities. Pokemon GO~\cite{GO} showed the exciting possibilities of overlaying a virtual space over a real one and how both the virtual and real space can support and enable entirely different interaction norms. Can we use such hybrid spaces to motivate counternormative behaviors that facilitate change? Two examples of prior work explore this direction. Inneract~\cite{Inneract} was an app that allowed users to create spaces with their own rules for interaction. For instance: ``we trade secrets in this space''. It allowed other users within a certain physical radius to ``enter'' spaces they were close to by opting into the interaction norms of those spaces. Similarly, Situationist~\cite{Situationist} was an app that allowed people to pick counternormative experiences they would like to be a part of (such as ``giving a 5-second hug'') and when they were in physical proximity to another user of the app, they were prompted to engage in the experience together. While both these project were short-lived social experiments, future work can further explore the design of experiences in alternate realities that foster social connection in teams and improve team climate.

\subsubsection{Generalizing to long-lived teams: identifying opportune moments to intervene}
While our study focused on short-lived teams, it illuminated important questions on designing interventions for well-being of long-lived teams. Our short-term study did not allow us to measure the persistence of \sys{}'s effects on perspective-taking but by investigating how effects of such team interventions dampen over time, future work can formulate strategies for periodic deployment of interventions that maintain social perceptiveness in teams. In our setting, \sys{} was triggered automatically and teams might well need such hand-holding in initial interactions~\cite{Whiting2020-pq, Zhou2018-mk}. However, once teams move past initial interactions and learn to work and engage with each other, should the intervention switch to being passive? This would allow team members to invoke the system when necessary (e.g. on arrival of a new member) as opposed to the system deciding the right time to interrupt teams. Future work can investigate stages in a team's interaction after which the system can step back and shift control to the users.

\section{Conclusion}
With \sys{}, we have demonstrated the promise of introducing spaces for reflection and perspective-taking as a means to foster open communication, encourage expression of diverse and conflicting viewpoints, and, ultimately, improve team satisfaction and viability among members of ad-hoc teams. This work provides a foundation for designing systems that leverage social perceptiveness to foster greater safety and efficacy within virtual teams. More broadly, we hope that, going forward, this work signals a greater focus on well-being and group climate as greater priorities for the design of technological supports for virtual groups.

\begin{acks}
We are grateful to Mark Whiting and Andrew Mao for their valuable inputs and guidance on conducting collaborative real-time experiments on Amazon Mechanical Turk. Our experimental platform was built atop TurkServer- a package Andrew helped develop and maintain- and we thank him for helping us understand and work with TurkServer. We also thank Yasmine Kotturi, Julia Cambre, Sayan Chaudhry, Franklin Li, Jaemarie Solyst, and our anonymous reviewers for their valuable comments and feedback on drafts of this work. We thank the participants in our pilots and the Amazon Mechanical Turk workers who participated in our study. This work was supported by the Office of Naval Research but this article solely reflects the opinions and conclusions of the authors and not our funder.
\end{acks}

\bibliographystyle{ACM-Reference-Format}
\bibliography{humanistic-management}

\end{document}